\documentclass[preprint]{aastex}
\usepackage{epsf}

\def\secpoint{\mbox{$''\mskip-7.6mu.\,$}}
\shorttitle{Chamaeleon I: Three Micron Survey}
\shortauthors{KENYON AND GOMEZ}
\begin{document}

\title{A Three Micron Survey of the Chamaeleon I Dark Cloud
\footnote{Based on observations obtained with the SPIREX/Abu system in
Antarctica. SPIREX/Abu operated for the 1999 observing season under
agreement between the National Optical Astronomy Observatories (NOAO) and
the Center for Astrophysical Research in Antarctica (CARA).}}

%% Use \author, \affil, and the \and command to format
%% author and affiliation information.
%% Note that \email has replaced the old \authoremail command
%% from AASTeX v4.0. You can use \email to mark an email address
%% anywhere in the paper, not just in the front matter.
%% As in the title, you can use \\ to force line breaks.

\author{Scott J. Kenyon}
\affil{Smithsonian Astrophysical Observatory, 60 Garden Street, Cambridge MA 02138}
\email{skenyon@cfa.harvard.edu}

\and

\author{Mercedes G\'omez}
\affil{Observatorio Astron\'omico de C\'ordoba, Laprida 854, 5000 C\'ordoba, Argentina}
\email{mercedes@oac.uncor.edu}

\affil{$  $}
\affil{\it to be published in the}
\affil{\bf Astronomical Journal}
\affil{\it May 2001}

%% Notice that each of these authors has alternate affiliations, which
%% are identified by the \altaffilmark after each name.  Specify alternate
%% affiliation information with \altaffiltext, with one command per each
%% affiliation.

%% Mark off your abstract in the ``abstract'' environment. In the manuscript
%% style, abstract will output a Received/Accepted line after the
%% title and affiliation information. No date will appear since the author
%% does not have this information. The dates will be filled in by the
%% editorial office after submission.

\begin{abstract}

We describe an $L$-band photometric survey of $\sim$ 0.5 deg$^2$ of the 
Cha I dark cloud.  The survey has a completeness limit of $L < 11.0$.
Our survey detects 124 sources, including all known pre-main sequence 
stars with $L$ $\le$ 11.  The fraction of sources with near-IR excess 
emission is 58\% $\pm$ 4\% for K = 9--11. Cha I sources have bluer 
$H-K$ and $K-L$ colors than pre-main sequence stars in Taurus-Auriga.  
These sources also have a strong correlation between EW(H$\alpha$) 
and $K-L$.  Stars with $K-L \le$ 0.6 have weak H$\alpha$ emission; 
stars with $K-L \ge$ 0.6 have strong H$\alpha$ emission.  Because
many Cha I sources are heavily reddened, this division between weak 
emission T Tauri stars and classical T Tauri stars occurs at a 
redder $K-L$ than in Taurus-Auriga.

\end{abstract}

%% Keywords should appear after the \end{abstract} command. The uncommented
%% example has been keyed in ApJ style. See the instructions to authors
%% for the journal to which you are submitting your paper to determine
%% what keyword punctuation is appropriate.


%% From the front matter, we move on to the body of the paper.
%% In the first two sections, notice the use of the natbib \citep
%% and \citet commands to identify citations.  The citations are
%% tied to the reference list via symbolic KEYs. The KEY corresponds
%% to the KEY in the \bibitem in the reference list below. We have
%% chosen the first three characters of the first author's name plus
%% the last two numeral of the year of publication as our KEY for
%% each reference.

\section{Introduction}

Pre-main sequence T Tauri stars (TTS) often display excess
infrared (IR) emission compared to a normal stellar photosphere.
The circumstellar material which produces this emission has
temperatures ranging from $\sim$ 50 K up to $\sim$ 2000 K 
\citep[e.g.,][]{ruc85,ryd87}.  Large disks surrounding TTS can produce 
these temperatures by reprocessing stellar radiation or by material 
flow through the disk at rates of 1--10 $\times ~ 10^{-8} ~ M_{\odot}$ 
yr$^{-1}$ \citep{ber88,ken87}.  
Disks can also produce the eruptions observed in pre-main sequence 
stars \citep[e.g.,][]{har96}.  After years of detailed 
comparisons between data and disk models for the energy distributions 
and other physical properties of T Tauri stars, recent images 
from {\it Hubble Space Telescope} \citep[e.g.,][]{bur96,ghe97,kri00} 
and ground-based radio interferometers \citep[e.g.,][]{koe93,wil96,wil00} 
dramatically confirmed that circumstellar disks are responsible for
the IR excesses of pre-main sequence stars.

Measuring the evolution of disks surrounding pre-main sequence
stars is important for understanding the evolution of the
central star and the evolution of the disk into a planetary
system \citep[e.g.,][]{har98,lad99}.  Large surveys 
of nearby star-forming regions are needed to derive the samples 
of pre-main sequence stars required for estimates of stellar ages
and disk properties \citep[e.g.,][]{kh95}.  Modern
instruments at optical, near-IR, mid-IR, and X-ray wavelengths 
can now provide such samples fairly routinely.

Here we report results of an $L$-band photometric survey of the
central portion of the Cha I dark cloud, an active star-forming
region with more than 150 pre-main sequence stars \citep[e.g.,][]{law96}.  
Our goal is to constrain properties of the disks in these pre-main sequence 
stars using the $K-L$ color index for a complete sample of stars 
selected at $K$ and $L$.  This survey complements our recent $JHK$ 
imaging survey \citep{gom01} and other near-IR surveys 
\citep{cam98, per99, oas99}.
The new $L$-band data also provide a needed link between near-IR
surveys and mid-IR and far-IR surveys with {\it ISO} and {\it IRAS}
data \citep[e.g.,][]{per00,bau84}. 

We describe the $L$-band survey in \S2, analyze the data in \S3, 
and conclude with a brief summary in \S4.

\section{Observations and Data Analysis}

We acquired $L$-band photometry of Cha I sources and standard stars 
using the South Pole Infrared Explorer (SPIREX) during the 1999 March 
to October observing season.  The SPIREX 60 cm telescope developed by
CARA for observations at the South Pole was equipped with the NOAO/KPNO
Abu infrared camera (Fowler et al. 1998), which includes a first 
generation 1024 $\times$ 1024 InSb detector array.  With a scale of
0\secpoint6 pixel$^{-1}$, the Abu field-of-view was $\sim$
10\arcmin~$\times$~10\arcmin~for these observations.
We divided the Cha I field into a 12 $\times$ 5 grid roughly coincident 
with our coverage from a $JHK$ survey using CIRIM at the Cerro Tololo
Interamerican Observatory \citep{gom01}.  Here we describe 
photometry derived from the central 100\arcmin~$\times$~20\arcmin~of 
the grid, acquired on 17 and 18 July 1999.  Registration problems 
prevented us from matching other $L$-band sources to $JHK$ sources from 
the CTIO survey.

J. Kastner and collaborators at the Rochester Institute of Technology (RIT)
performed the initial data reduction of the SPIREX/Abu images. They
developed a library of IDL processing algorithms to flat-field and
to register five dithered images per target field.  They also derived
sky conditions each night, using measured backgrounds and standard stars.
Our observations were generally made in good conditions with below-average
backgrounds\footnote{A detailed description can be found at:
http://pipe.cis.rit.edu/index.html}.

We derived calibrated $L$-band photometry of Cha I sources using routines
in IRAF\footnote{IRAF is distributed by the National Optical
Astronomy Observatory, which is operated by the Association of
Universities for Research in Astronomy, Inc. under contract to the
National Science Foundation.}.  We used IRAF PHOT with an 8 pixel 
aperture to derive the standard star calibration and photometry of 
Cha I sources.  This aperture encircles 95\% or more of the light of
all sources, based on measurements of the point-spread function for 
standard stars and bright Cha I sources.  We used measured backgrounds
and observations of 8--9 standard stars on 17--21 July 1999 to establish 
the extinction correction for the Earth's atmosphere and the photometric 
zero-point. We adopted $L$-band magnitudes for the standards from the 
SPIREX/Abu web page cited previously.  The $L$-band 
calibration and extinction correction are accurate to $\pm$10\% 
or better.  Comparisons of our Cha I photometry with previous 
measurements suggest a typical error of $\pm$0.05 mag for bright 
sources with $L$ $=$ 7--8, $\pm$0.10 mag for sources with $L$ $=$ 9--11, and
$\pm$0.20 mag for sources at the detection limit, $L$ $\ge$ 11.

Table 1 lists $L$ magnitudes and $K-L$ colors for 124 $L$-band sources detected
in our survey.  The J2000.0 coordinates and the $JHK$ data in Table 1 are 
from the CIRIM survey. To make reliable matches for the $L$-band sources,
we derived preliminary coordinates from the $L$-band images using the
WCSTools routines\footnote{Available at ftp://cfa-ftp.harvard.edu/pub/gsc/WCSTools}. 
We adopted the nominal pixel scale and chose a rough center for each frame 
by eye.  We made one iteration of the center to minimize the average deviation
between $JHK$ and $L$ coordinates for the 3--6 sources with $L$ $\le$ 10 in 
each image, and then matched fainter sources one-by-one.  The small
number of sources per image prevented us from making a more accurate
fit for the plate solution on most images.

We made several tests to estimate the completeness limit for the 
SPIREX/Abu survey.  Figure 1 compares counts in half magnitude bins 
(left panel) with number counts for the entire population of known
pre-main sequence stars in Cha I (right panel).  Both sets of counts 
rise monotonically to $L$ $=$ 11 and then turn over.  These 
results suggest a completeness limit of $L$ $=$ $10.75 \pm 0.25$. To
confirm this estimate, we used $K$ and $H-K$ from our CIRIM survey
of Cha I \citep{gom01} to predict $L$ magnitudes assuming
$K-L \approx$ $H-K$ $\pm$ $\delta m$.  We tried many Monte Carlo 
trials for random $\delta m$ in the range 0.1--0.4 mag.  We detected 
all CIRIM sources with $K <$ 10.5 in our survey region.  For fainter 
sources, this test predicts 10 undetected $L$-band sources with 
$K <$ 12 independent of the adopted error, $\delta m$.
Two of these, ISO 103 and 110826$-$765559 from \citet{gom01}, 
have $K$ = 10.5--11.0 
and $H-K <$ 0.5; the rest have $K > 11$ and $H-K < $ 1 and should 
lie close to our detection limit of $L$ = 11.  Thus, our survey is 
essentially complete for stars with $L <$ 11, which are comparable in 
brightness to the faintest known pre-main sequence stars in the cloud.

\section{Analysis}

Our $L$-band survey of 124 sources covers a modest fraction, $\sim$ 
0.5 deg$^2$, of the Cha I star-forming region.  We recover all 
previously known near-IR sources in the region, including 15 
discovered in our deep $JHK$ survey with CIRIM at CTIO \citep{gom01}.
 Only $\sim$ 40 stars in this sample have
reliable H$\alpha$ equivalent width measurements; fewer than
half of the sample have reliable spectral types.  A complete 
analysis thus requires additional optical spectra for the entire
sample, which is beyond the scope of this paper. Here, we consider
the near-IR color-color diagrams (\S3.1), the fraction of sources
with near-IR excesses (\S3.2), and a correlation of two indicators
of disk accretion, $K-L$ and the equivalent width of H$\alpha$
(\S3.3).

\subsection{Color-color diagrams}

Figure 2 shows, side by side, the $J-H$ vs $H-K$ and $J-H$ vs $K-L$ 
color-color diagrams for all previously known young stellar objects in 
the Cha I cloud. We compiled from the literature $L$ magnitudes for known 
pre-main sequence stars missed by our $L$-band survey (due mainly to frame 
registration problems) or saturated in our images
\citep{gla79,gau92,pru92,law96}.
We used the online Catalog of Infrared Observations\footnote{Available at 
http://ircatalog.gsfc.nasa.gov} to check the completeness of this sample.
The dot-dashed lines indicate the expected locus of reddened stars 
with the intrinsic colors of normal main sequence stars \citep{bes88,gom01}.
Stars lying outside this band have near-IR excesses from circumstellar dust
\citep{all72,hyl80,hyl82}.  The $J-H$ vs $K-L$ 
diagram shows more stars 
with near-IR excess than the $J-H$ vs $H-K$ diagram. The 3 $\mu$m data 
are less compromised by interstellar reddening and provide more contrast 
relative to photospheric emission from the central star than shorter 
wavelength observations \citep[see Haisch, Lada, \& Lada 2000;]
[and references therein]{lad00,kh95} .

Figure 3 compares the frequency distribution of the $K-L$ color index in 
Cha I with the distribution in the Taurus-Auriga dark cloud \citep{kh95}.
We show two distributions of Cha I stars, our SPIREX/Abu sample and a 
combined sample using data compiled from the literature to supplement 
SPIREX/Abu data.  These two distributions have the same median $K-L$, 0.57 
(SPIREX/Abu sample) and 0.58 (combined sample), and a probability of more 
than 99.9\% of being drawn from the same parent distribution according
to a K-S test. The Cha I frequency distribution consists of a single peak 
at $K-L \approx$ 0.4--1.6 and a red tail which extends to $K-L \approx$
2.2.  The small peaks at $K-L \approx$ 1.1 and $K-L \approx$ 1.5 are due
to small number statistics according to a K-S test.  The Taurus-Auriga 
distribution has two peaks, weak emission T Tauri stars (WTTs) with 
$K-L \le$ 0.2 and classical T Tauri stars (CTTs) with $K-L \approx$ 
0.8-1.0 \citep{kh95}.  The CTTs peak in Taurus-Auriga has a broad 
shoulder extending to $K-L \approx$ 3 which includes many class I protostars.  

Two main differences between the Cha I and Taurus-Auriga pre-main sequence 
stars produce the different $K-L$ frequency distributions.  The Taurus-Auriga
cloud has many more protostars with $J-H \ge$ 3 and $H-K$ $\approx$ 
$K-L \ge$ 2 than does Cha I.  The Cha I population also does not show 
a peak at $K-L \le$ 0.2,  because WTTs in Cha I are more heavily reddened 
than WTTs in Taurus-Auriga.  For stars with accurate spectral types,
the median optical reddening is $A_V \sim$ 0.6 mag for Taurus-Auriga WTTs 
\citep{kh95} and $A_V \sim$ 1.6 mag for Cha I WTTs 
\citep[e.g.,][]{law96}.
Our near-IR color-color diagrams suggest a population of fainter and much 
more heavily reddened Cha I WTTs with $A_V \sim$ 5--10.  Optical or 
near-IR spectra of these sources are necessary to test this prediction.

Despite the larger reddening, the median near-IR colors of young stars
in Cha I are bluer than the median near-IR colors of young stars in
Taurus-Auriga.  We derive a median $K-L$ of 0.58$^{+0.35}_{-0.16}$ for  
Cha I sources and 0.73$^{+0.33}_{-0.48}$ for Taurus-Auriga stars. 
The error bars indicate the inter-quartile ranges for each sample.
The difference in the observed $K-L$ color has a 2$\sigma$ significance
according to a K-S test.  Estimating the intrinsic color difference 
between Cha I and Tau-Aur sources is difficult, because most of the Cha I
stars do not have measured spectral types.  Adopting K0-M6 spectral 
types for these stars, we can estimate $A_V$ from the observed near-IR
colors and make a crude comparison with Tau-Aur sources with measured
$A_V$.  This exercise increases the significance of the color
difference to 3--4$\sigma$, depending on the very uncertain reddening
correction.  Acquiring good spectral types for these Cha I sources
would allow a better test.

The $K-L$ frequency distribution in Cha I has some features in common
with the frequency distribution in $\rho$ Oph (Figure 3).  We compiled
$JHK$ data from \citet{bar97} and $L$ data from the Catalog of Infrared
Observations \citep[see][and references therein]{el78,wil83,com93}.
Both clouds
have a single, well-defined peak at $K-L <$ 1.  In $\rho$ Oph, this
peak consists of heavily reddened WTTs and a few moderately reddened
CTTs.  Unlike Cha I, $\rho$ Oph has a population of very red protostars
as in the Taurus-Auriga cloud.  The second peak in the $\rho$ Oph 
distribution at $K-L \approx$ 2.0-2.5 has a 2-3$\sigma$ significance 
according to a K-S test.  Some of the stars in this peak are heavily 
reddened CTTs.

\subsection{Near-IR excesses}

In the current picture of star formation, young stars evolve from deeply 
embedded protostars into CTTs and then into WTTs before arriving on the 
main sequence.  The CTTs are pre-main sequence stars surrounded by a 
circumstellar disk; the flow of material through the disk onto the star 
produces an optical/ultraviolet excess and an IR excess over photospheric
emission from the central star.  Reprocessing of stellar radiation by
the disk also produces some of the IR excess \citep[e.g.,][]{ken87,chi97} .
The WTTs do not have excess emission,
either because the disk has dispersed or because the disk emits too
little radiation to produce an excess.  

The `disk fraction' is an interesting parameter in this picture.  The
disk fraction is the fraction of T Tauri stars with IR excess emission.
Because long wavelength near-IR colors provide more contrast relative
to emission from the stellar photosphere, recent studies have used 
the $K-L$ color to measure the disk fraction \citep[e.g.,][]{hai00}.
Longer wavelength colors such as $K-N$ or $K-Q$ would 
probably provide a better measure \citep[e.g.,][]{kh95}, 
but these data are difficult to obtain for large samples of pre-main
sequence stars with current technology.

To estimate the number of sources with $K-L$ excesses in Cha I, we employ 
a Monte Carlo simulation to account for uncertainties in the $K-L$ colors
and the uncertainty in the slope of the reddening band. To construct 
the simulation, we begin with ($J-H$,$K-L$) pairs for our sample,
($J-H$,$K-L$) colors of normal main sequence stars from \cite{kh95},
and an adopted slope for the reddening law,
$E_{J-H}/E_{K-L}$ = 1.88 from \cite{bes88}.
We adopt an error of $\pm$ 0.05 in the slope of the reddening law
\citep[see][]{bes88,ken98,gom01}.
For each Monte Carlo trial, we replace 
each color index with an artificial color index $c_i^{\prime}$,

\begin{equation}
c_i^{\prime} = c_i + e_i g_i ~ ,
\end{equation}
 
\noindent
where $c_i$ is the observed color index, 
$e_i$ is the error in the color index, and
$g_i$ is a normally distributed deviate with zero mean
and unit variance \citep{pre92}. The reddening law for each trial is

\begin{equation} 
E_{J-H}/E_{K-L} = 1.88 + 0.05 g_i ~ , 
\end{equation} 
  
\noindent
We then derive the fraction of sources which lie to the right of the
reddening band in the color-color diagram.  After 10,000 such
trials, we evaluate the median fraction of IR excess sources and
the inter-quartile range of the distribution.  This simulation
allows us to estimate the error in the fraction of IR excess sources
and to evaluate the sensitivity of this fraction to uncertainties in
the near-IR colors, the intrinsic main sequence star colors, and the 
slope of the reddening band.

Our set of Monte Carlo trials results in an observed fraction of
sources with $K-L$ excess emission, 58\% $\pm$ 4\%. The 1$\sigma$
uncertainty includes the uncertainty in the observed colors and 
the slope of the reddening law.  This fraction does not depend on
errors in the intrinsic colors of main sequence stars used to define
the reddening band.  The fraction of IR excess sources is sensitive
to the real slope of the reddening band.  In general, there are fewer
near-IR excess sources if the slope of the reddening law is shallower
than the adopted law.  In practice, most Cha I sources with $K-L$ excesses
lie well away from the reddening band.  Unless the real slope of the
reddening law is 3$\sigma$ or more shallower than the adopted slope,
this uncertainty does not affect our result significantly.

The disk fraction of 58\% $\pm$ 4\% in Cha I is smaller than disk
fractions estimated for Taurus-Auriga, 69\%, the Trapezium cluster,
80\% $\pm$ 7\% \citep{lad00}, and NGC 2024, 86\% $\pm$ 8\% \citep{hai00}.  
Cha I has fewer class I protostars than does Taurus-Auriga \citep{kh95};
Cha I probably has fewer protostars than NGC 2024 and the Trapezium cluster.
We suspect that Cha I has fewer protostars and a smaller disk fraction, 
because it contains a larger fraction of older pre-main sequence
stars than Taurus-Auriga, the Trapezium cluster, or NGC 2024 
\citep{kh95,law96,har98,hai00,lad00}.  Measurement of a smaller disk 
fraction, 65\% $\pm$ 8\%, in the older cluster IC 348 supports this
conclusion \citep{hai01}. 
Age estimates for newly discovered pre-main sequence stars in
Cha I and comparisons with other star-forming regions are needed
to confirm relationships between stellar age and disk fraction.

To examine the disk fraction and IR excesses of Cha I sources in 
more detail, we considered the $K-m(6.7)$ and $K-m(14.3)$ colors
derived from our data and data from the ISO mission \citep{nor96,per00}.  
With the calibration of \citet{olo99}, we constructed
color-color diagrams using our $J-H$ and $K-L$ colors to search for 
correlations between near-IR and mid-IR excesses.  As expected,
WTTs have bluer $K-m(6.7)$ and $K-m(14.3)$ colors than CTTs. 
We found weak (1--2$\sigma$) evidence for a deficit of Cha I
sources with $K-m(14.3)$ $\approx$ 3--4.  The Taurus-Auriga cloud 
has a considerable deficit of sources with $K-N \approx$ 0.75--2.25, 
producing a pronounced gap between the CTTs and the WTTs in this cloud
\citep[see][and references therein]{kh95}.  Although this deficit
is interesting, the Cha I sample is too small and too heavily reddened
to make a strong case for a deficit similar to the one observed in
Taurus-Auriga.  Future mid-IR observations with the {Space Telescope 
Infrared Facility} and other instruments will yield better statistics
on the distribution of mid-IR excesses in Cha I sources.

\subsection{H$\alpha$ Equivalent Widths}

In most circumstellar disk models, the disk emits in the IR while 
an accretion stream or boundary layer produces optical continuum 
emission and some low ionization emission lines 
\citep[e.g.,][]{lyn74,ken87,ber88,ken96}.
In this picture, near-IR and optical excess emission should be
correlated.  For Taurus-Auriga pre-main sequence stars, there are
strong correlations between IR excess emission and the optical
continuum and emission lines \citep[e.g.,][]{hart90,hart93,bas90,edw93,
val93,kh95}.  

To test for a correlation between optical and IR excess emission in 
Cha I sources, we collected H$\alpha$ equivalent widths, EW(H$\alpha$), 
from the literature 
\citep{heme73,app83,wal92,gau92,hart93,hue94,alc95,law96,com99}.
We took the average for sources with two or more measurements;
using the median makes no difference in our conclusions.

Figure 4 shows the relationship between EW(H$\alpha$) and $K-L$
for 40 sources in the SPIREX/Abu survey region with good 
measurements.  There is a strong correlation; the Spearman 
rank correlation coefficient is $r_s$ = 0.70. The probability that 
the observed distribution is drawn from a random distribution with 
no correlation is $p_D \approx 6 \times 10^{-7}$.
As in Taurus-Auriga, all of the power in this correlation results
from differences between WTTs and CTTs.  The probability that the
observed distribution is random increases to
$p_D \approx 4 \times 10^{-5}$ for sources with $K-L \ge$ 0.4 and
$p_D \approx$ 0.13 for sources with $K-L \ge$ 0.6.  Because some
Cha I WTTs are heavily reddened, the division between WTTs and CTTs 
in Cha I occurs at a somewhat larger $K-L$ than in Tau-Aur 
\citep[see][]{kh95}.

As far as we know, there are insufficient measurements of $U-B$ colors or 
of optical or ultraviolet veiling in Cha I to attempt correlations with the
IR excess emission.  Nevertheless, the strong correlation between
EW(H$\alpha$) and $K-L$ indicates that accretion powers the 
optical emission lines in Cha I CTTs.

\section{Summary}

We have described an $L$-band photometric survey of part of the 
Cha I dark cloud.  We detect 124 sources with $L < 11.0$,
the approximate completeness limit.  Our survey recovers all 
known pre-main sequence stars in the region and yields new 
measurements of near-IR excess emission in 15 candidate pre-main
sequence stars from our deep $JHK$ survey \citep{gom01}.
The fraction of sources with near-IR excess emission is 
58\% $\pm$ 4\%, based on a Monte Carlo simulation of the 
observations using an adopted reddening law.  This fraction 
is constant for sources with K = 9--11. 

Cha I sources have bluer $K-L$ colors than pre-main sequence stars 
in Taurus-Auriga.  The $H-K$ colors are also bluer, despite the 
larger apparent reddening of some Cha I WTTs.  Optical spectra
are needed to derive spectral types and estimate the reddening for
these sources.  Accurate values for the optical extinction $A_V$
would provide a better test of the difference between the near-IR
colors for Cha I pre-main sequence stars and young stars in other
clouds.

We find a strong correlation between EW(H$\alpha$) and $K-L$ for
$\sim$ 40 Cha I sources.  Stars with $K-L \le$ 0.6 have weak
H$\alpha$ emission; stars with $K-L \ge$ 0.6 have strong H$\alpha$
emission.  This result agrees with predictions of accretion disk 
models, where a disk produces the near-IR excess and an accretion
stream or boundary layer produces an optical/ultraviolet continuum
and strong emission lines.  There is weak evidence that the median
EW(H$\alpha$) is smaller for these Cha I sources than for pre-main
sequence stars in the Taurus-Auriga cloud.  Magnetic accretion disk
models, where the stellar magnetic field truncates the disk above
the stellar photosphere, predict a smaller median EW(H$\alpha$) 
for sources with bluer $K-L$ colors \citep{ken96}. 
If these trends for Cha I sources are correct, these data provide 
weak evidence for smaller accretion rates among Cha I CTTs than 
among Taurus-Auriga CTTs.  Optical spectra for the remaining 80 stars 
in this Cha I field would yield a definitive test of these trends.

\acknowledgments
 
We are grateful to the SPIREX/Abu staff, especially C. Kaminski and 
K. M. Merrill, for acquiring the observations and for assistance with
planning a successful observing strategy.  A. Fowler and N. Sharpe
from NOAO modified the Abu system for use at the South Pole and 
made it operational on site.
We also thank J. Kastner and collaborators for performing the initial
data reduction of the SPIREX/Abu images.  D. Mink kindly assisted us
with the WSCTools software.  
We thank J. Kastner, C. Lada, W. Lawson (the referee), and K. M. Merrill
for helpful comments and questions.

\clearpage

%% No more than seven \figcaption commands are allowed per page,
%% so if you have more than seven captions, insert a \clearpage
%% after every seventh one.

%% There must be a \figcaption command for each legend. Key the text of the
%% legend and the optional \label in curly braces. If you wish, you may
%% include the name of the corresponding figure file in square brackets.
%% The label is for identification purposes only. It will not insert the
%% figures themselves into the document.
%% If you want to include your art in the paper, use \plotone.
%% Refer to the on-line documentation for details.

%\centerline{Figure Captions}

\epsffile{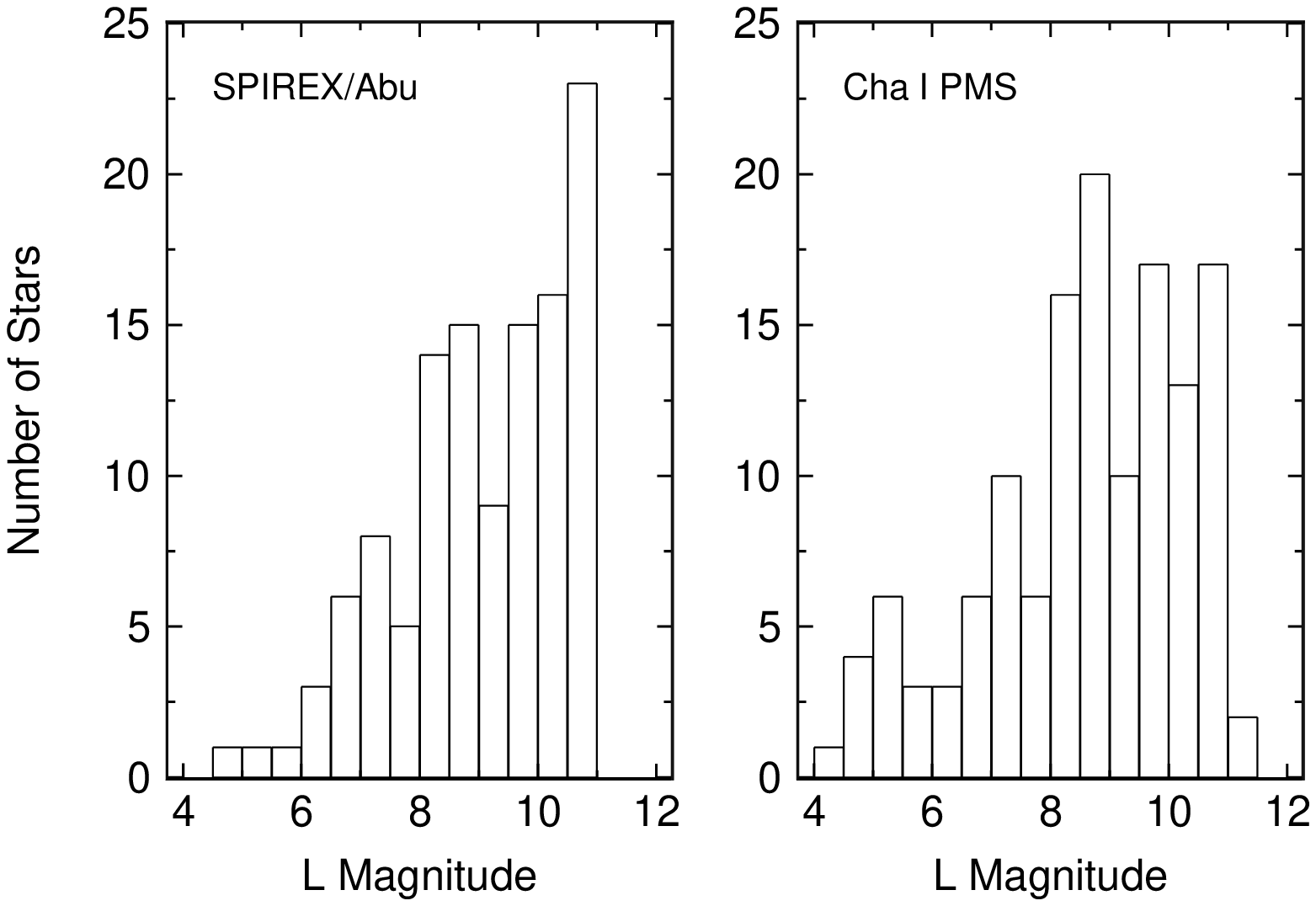}
\figcaption[Kenyon.fig1.eps]{Number counts of Cha I sources.
Left panel: SPIREX/Abu sources from this paper.
Right panel: Known pre-main sequence stars in Cha I.}

\epsffile{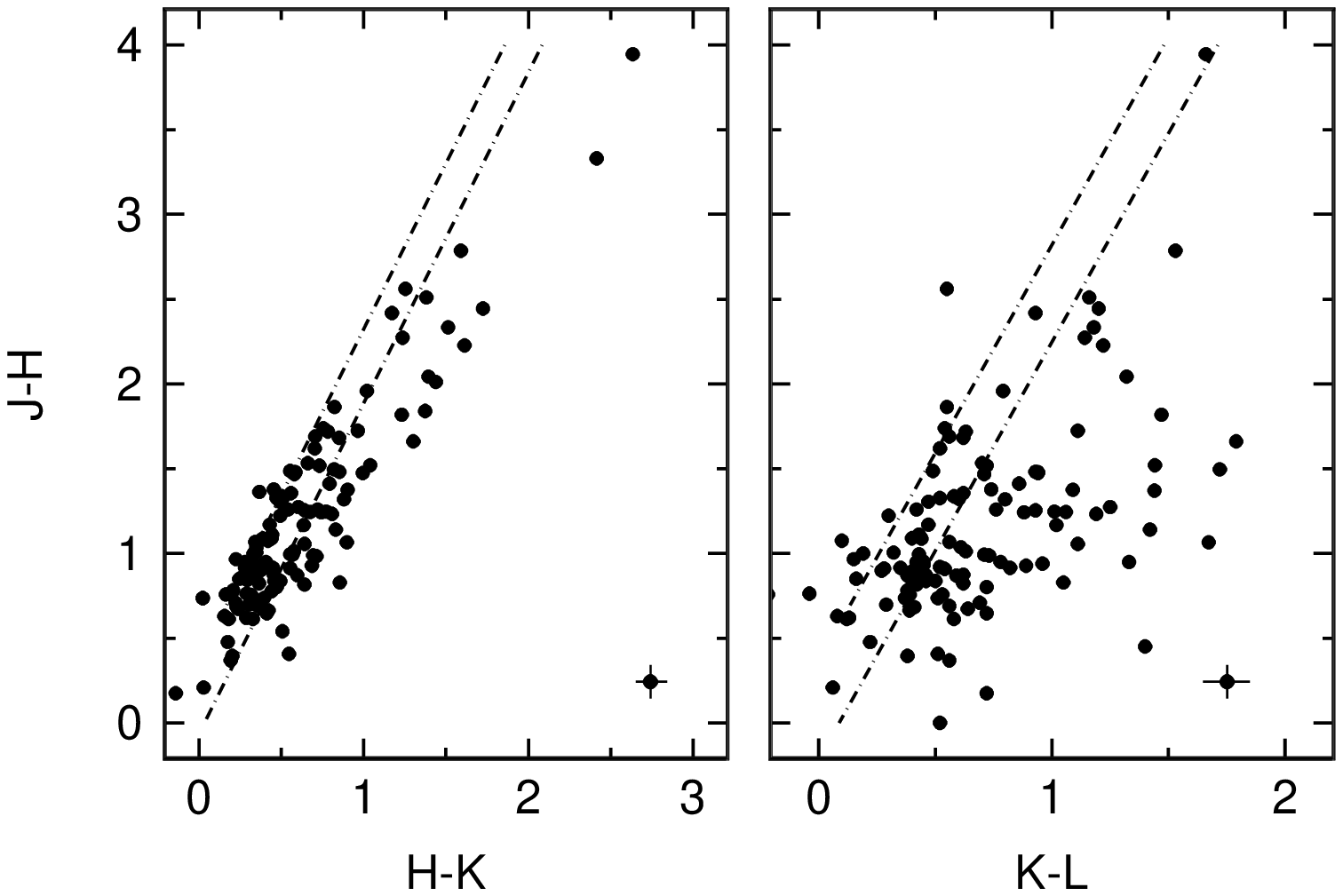}
\figcaption[Kenyon.fig2.eps]{Near-infrared color-color diagrams
for SPIREX/Abu survey.
Left panel: $J-H$ vs $H-K$.
Right panel: $J-H$ vs $K-L$.
The dot-dashed lines in each panel bracket the expected locus of
reddened colors of main sequence stars \citep{bes88}.
The error bar in the right corner of each panel indicates 1$\sigma$
errors for each color.}

\epsfxsize=9.0in
\epsffile{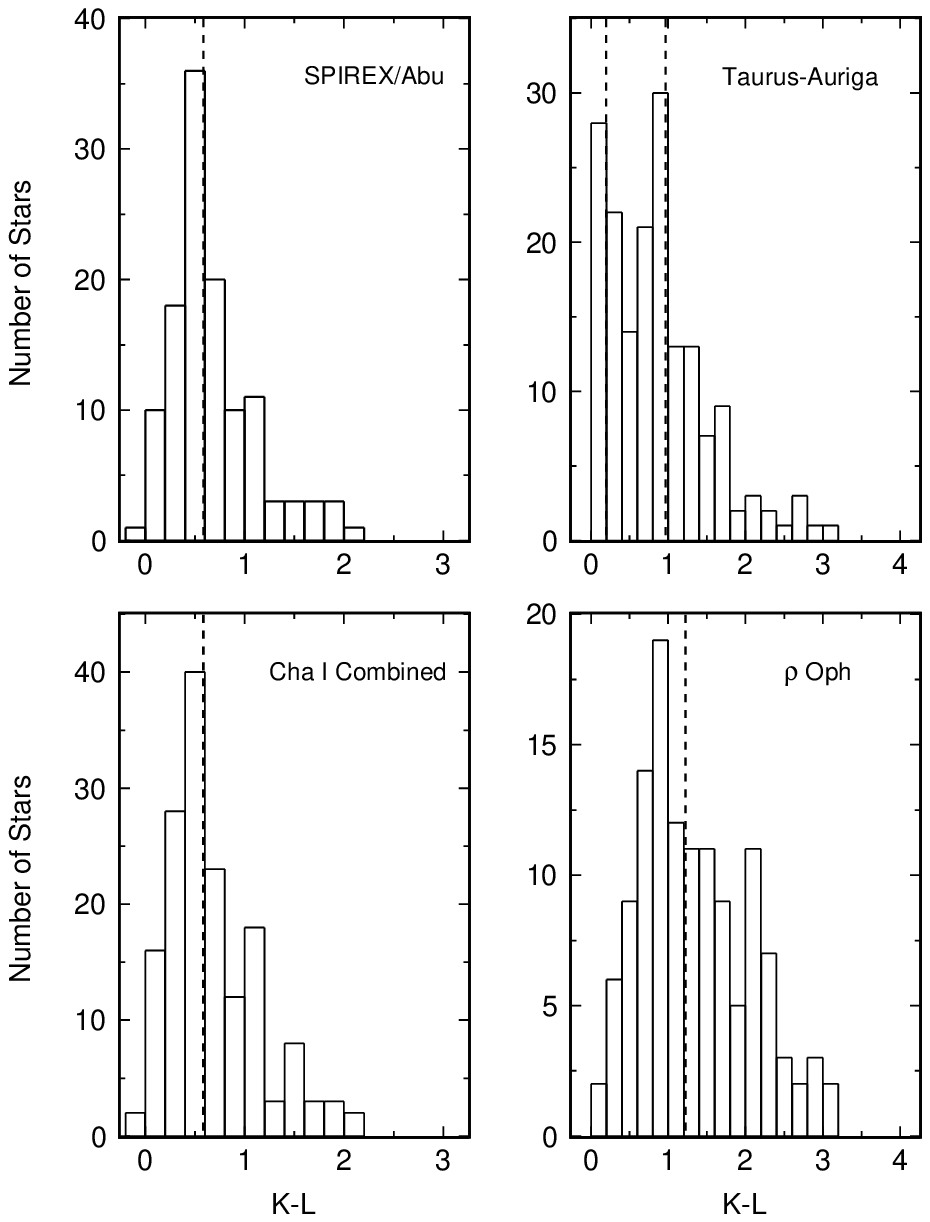}
\figcaption[Kenyon.fig3.eps]{Frequency histograms for the $K-L$ color
in the Cha I, $\rho$ Oph, and Taurus-Auriga dark clouds.  
Upper left panel: Cha I SPIREX/Abu sample.
Lower left Panel: Cha I combined sample.
Upper right panel: Taurus-Auriga.
Lower right Panel: $\rho$ Oph, several $\rho$ Oph stars with 
$K-L \ge$ 4 have been excluded for clarity.
The dashed lines indicate the median $K-L$ for the
$\rho$ Oph and both Cha I samples. In Taurus-Auriga,
the dashed lines mark the median $K-L$ colors for
WTTs (left dashed line) and for CTTs (right dashed line).}

\epsffile{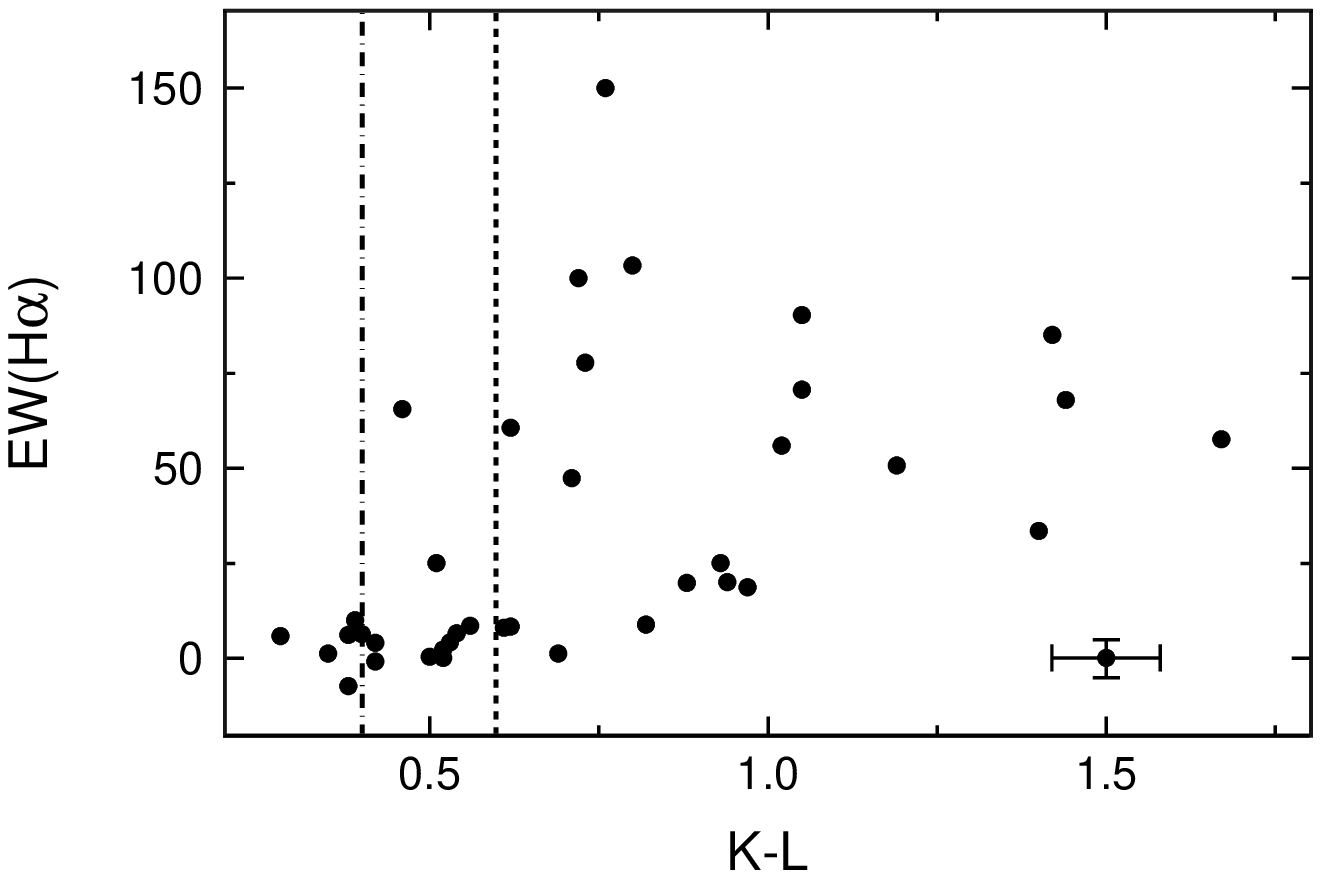}
\figcaption[Kenyon.fig4.eps]{H$\alpha$ equivalent widths as a function
of $K-L$ color for SPIREX/Abu sources. The dot-dashed line indicates the division at $K-L$ = 0.4
between WTTs and CTTs in Taurus-Auriga.  The dashed line indicates the
division at $K-L$ = 0.6 between reddened WTTs and CTTs in Cha I.}

\newpage

\begin{deluxetable}{rccrcccll}
\tabletypesize{\scriptsize}
\tablecaption{$L$ Band Observations for the Cha I sources}
\label{tbl-1}
\tablewidth{0pt}
\tablehead{  
\colhead{No.} & \colhead{$\alpha$(2000.0)} & \colhead{$\delta$(2000.0)} &
\colhead{$L$} & \colhead{$K-L$} & \colhead{$H-K$} & \colhead{$J-H$} &
\colhead{Id.} & \colhead{Ref.} }
\startdata                                     
1& 11 07 11.0 & $-$77 46 41 &   9.11 & 0.61 & 0.35 &  1.13 &	 Hn 6, ISO 94 &	7,12\\
2& 11 07 12.8 & $-$77 43 51 &   9.70 & 0.42 &	0.54 &  1.37 &   CHXR 22E &	4\\ 
3& 11 07 16.5 & $-$77 23 08 &   9.31 & 1.66 &	2.63 &	4.19 &   Denis 23, ISO 97 &	2,12 \\	
4& 11 07 16.9 & $-$77 26 22 &   sat  &   	  & 0.37 &	1.48 &   ISO 96	 &	12	  \\ 
5& 11 07 19.4 & $-$77 29 39 &   9.68 & 0.64 &	0.24 &	0.75 &   ISO 99  &  12         \\
6& 11 07 21.6 & $-$77 38 07 &   sat  &  	  & 0.44 &	0.86 &   Lh$\alpha$ 332-17, ISO 100 & 13, 12\\	
7& 11 07 21.7 & $-$77 22 12 &   9.30 & 1.53 &	1.59 &	2.97 &   Denis 24, B 35, ISO 101 & 2, 1, 12	\\
8& 11 07 23.7 & $-$77 41 25 &   6.90 & 0.74 &	0.45 &	1.49 &   Denis 25, ISO 102 & 2, 12 \\
9& 11 07 24.1 & $-$77 42 27 &  10.27 & 0.44 & 0.39 &	1.19 &  &				 \\     
10& 11 07 24.2 & $-$77 43 50 &  10.72 & 1.11 &	0.64 &	1.15 &  &			     \\ 
11& 11 07 26.3 & $-$77 01 52 &  10.95 & $-$0.22~~ & 	0.76 &  & & 				  \\    
12& 11 07 28.6 & $-$76 52 12 &   9.11 & 0.39 & 0.32 &	0.84 &  &			     \\ 
13& 11 07 32.3 & $-$77 28 25 &  10.18 & 0.69 &	0.22 &	0.79 &     CHXR 25, ISO 105 & 4, 12\\		
14& 11 07 33.7 & $-$77 11 07 &   9.44 & 0.38 &	0.21 &	0.86 &  & 		         \\
15& 11 07 35.3 & $-$77 34 51 &  10.55 & 0.40 &	0.34 &	0.98 &     B 34, CHXR 76, ISO 106 & 1, 4, 12\\	
16& 11 07 36.9 & $-$77 35 19 &   9.29 & 0.62 &	0.85 &	1.81 &     Denis 26, ISO 107 & 2, 12\\		
17& 11 07 37.0 & $-$77 33 33 &   8.52 & 0.70 &	0.66 &	1.65 &     CHXR 26, ISO 108  & 4, 12 \\	
18& 11 07 37.5 & $-$77 33 09 &  10.36 & 0.55 &	0.82 &	2.00 &     Denis 27, ISO 109 & 2, 12 \\		
19& 11 07 38.4 & $-$77 47 19 &  10.55 & 0.59 &	0.33 &	0.96 &     ISO 110 & 12		\\	    
20& 11 07 42.7 & $-$77 34 00 &   9.94 & 0.71 &	0.57 &	1.09 &     H$\alpha$2, ISO 111 &13, 12	\\	    
21& 11 07 44.5 & $-$77 39 43 &   7.40 & 1.02 &	0.63 &	1.27 &     HM 15, ISO 112  & 8, 12\\    
22& 11 07 46.1 & $-$77 38 05 &   7.33 & 0.93 &	1.17 &  2.59 &     Denis 28, ISO 113 & 2, 12\\		
23& 11 07 53.2 & $-$77 36 56 &  10.88 & 0.39 &	0.42 &	0.74 &     H$\alpha$3, ISO 116	& 3, 12\\	    
24& 11 07 56.7 & $-$77 27 27 &   7.29 & 0.52 &	0.30 &	1.01 &     CHX 10a,  ISO 117 & 5, 12\\		
25& 11 07 57.3 & $-$77 41 56 &   8.69 & 0.60 &	0.51 &	1.43 &     ISO 118	& 12\\		    
26& 11 07 57.5 & $-$77 17 27 &   8.29 & 1.32 &	1.39 &	2.19 &     B 38	& 1\\		    
27& 11 07 58.4 & $-$77 42 42 &   8.62 & 1.19 &	0.81 &	1.34 &     Sz 23, ISO 120 & 13, 12\\		    
28& 11 07 59.1 & $-$77 38 46 &   sat  &  	  & 1.04 &	1.64 &     HM 16, ISO 119 & 8, 12 	\\	
29& 11 08 00.2 & $-$77 15 33 &  10.38 & 0.72 &	0.47 &	0.89 &     ISO 121 & 12	\\		
30& 11 08 00.3 & $-$77 17 32 &   8.45 & 0.63 &	0.78 &	1.85 &     CHXR 30, ISO 122	& 4, 12 \\	
31& 11 08 01.9 & $-$77 42 29 &   sat  &  	  & 0.55 &	1.00 &     VW Cha, ISO 123	& 13, 12 \\	
32& 11 08 03.2 & $-$77 48 37 &  10.08 & 0.52 &	0.47 &	1.44 &     ISO 125	& 	12 \\
33& 11 08 04.2 & $-$77 38 43 &   sat  &   	  & 1.37 &	1.98 &     Denis 32, ISO 126 & 2, 12 \\	
34& 11 08 04.2 & $-$77 39 19 &   sat  &  	  & 0.51 &	0.61 &     HD 97048, ISO 124 & 12\\	
35& 11 08 10.1 & $-$76 42 26 &   8.67 & 0.06 &	0.03 &	0.26 &     ISO 129 &	12 \\	
36& 11 08 12.1 & $-$77 18 54 &   9.92 & 0.55 &	1.25 &	2.74 &     Denis 33, ISO 130 & 2, 12\\	
37& 11 08 12.9 & $-$77 19 13 &   7.74 & 1.16 &	1.38 &	2.68 &     Denis 34, ISO 131 & 2, 12 \\	
38& 11 08 15.1 & $-$76 37 33 &   9.94 & 0.56 &	0.19 &	0.43 &     ISO 133 & 12		\\	
39& 11 08 16.4 & $-$77 33 53 &   sat  &      & 0.71 &	1.08 &     Glass I, ISO 135 & 6, 12\\		
40& 11 08 16.9 & $-$77 44 37 &   9.71 & 0.42 &	0.28 &	0.97 &     HM 19, ISO 136	& 8, 12\\	
41& 11 08 17.4 & $-$77 44 11 &  10.19 & 0.51 &	0.39 &	0.82 &     ISO 137 & 12 \\		
42& 11 08 20.6 & $-$77 39 19 &  10.52 & 0.56 &	0.36 &	0.77 &     H$\alpha$4 &	3\\	    
43& 11 08 20.7 & $-$76 56 55 &   9.97 & 0.27 &	0.37 &	0.99 &     ISO 139 & 12	\\		
44& 11 08 20.7 & $-$77 05 22 &  10.71 & 0.15 &	0.23 &	1.06 &     ISO 140	& 12 \\		
45& 11 08 22.6 & $-$77 30 28 &  10.67 & 0.43 &	0.55 &	1.09 &     ISO 143 & 12	\\		
46& 11 08 23.9 & $-$76 24 06 &  10.50 & 0.08 &	0.16 &	0.70 &            &  \\
47& 11 08 24.7 & $-$77 41 47 &   9.94 & 0.82 &	0.43 &	1.01 &     H$\alpha$5, ISO 144	& 3, 12\\	   
48& 11 08 25.8 & $-$76 48 33 &   6.69 &      &      &  0.82 & 				  &  \\   
49& 11 08 38.7 & $-$77 43 51 &   6.60 & 1.47 & 1.23 &	1.96 &     IR-Neb, ISO 150 & 14, 12	\\	  
50& 11 08 39.2 & $-$77 16 05 &   8.22 & 0.80 &	0.88 &	1.43 &     HM 20, ISO 151 & 8, 12	\\	  
51& 11 08 39.6 & $-$77 34 17 &  10.58 & 0.46 &	0.49 &	0.92 &     H$\alpha$6, ISO 152	& 3, 12\\	 
52& 11 08 41.1 & $-$76 36 08 &   8.92 & 0.28 &	0.36 &	1.00 &     CHX 13a, ISO 153 & 5, 12\\		
53& 11 08 51.5 & $-$76 28 21 &  11.12 & 0.13 &	0.29 &	0.70 &     ISO 158	& 12 \\		    
54& 11 08 51.5 & $-$76 25 12 &  10.53 & 0.72 &	0.41 &	0.72 &     Sz 28, ISO 157 & 13, 12\\		
55& 11 08 54.3 & $-$76 49 29 &  11.13 & 0.41 &	0.23 &	0.76 &     ISO 159 & 12 \\	    
56& 11 08 54.6 & $-$77 32 12 &  10.55 & 0.62 &	0.36 &	0.91 &     CHXR 78C, ISO 160 & 4, 12	\\	
57& 11 08 54.7 & $-$76 51 29 &  10.04 & 0.16 &	0.24 &	0.94 &     ISO 163 & 12 \\			    
58& 11 08 55.1 & $-$77 02 13 &   8.13 & 1.42 &	0.83 &	1.24 &     VY Cha, ISO 162 & 13, 12\\		
59& 11 08 55.4 & $-$76 32 43 &  10.90 & 0.46 &	0.60 &	0.96 &     ISO 165 & 12\\			
60& 11 08 55.4 & $-$76 47 44 &  10.96 & 0.29 &	0.30 &	0.78 &     ISO 161 & 12\\			    
61& 11 08 55.6 & $-$77 04 52 &   7.04 & 0.56 &	0.34 &	1.17 &     ISO 166 & 12\\			    
62& 11 08 56.2 & $-$77 27 12 &  10.51 & 1.06 &	0.67 &	1.35 &     ISO 167 & 12\\			    
63& 11 08 56.6 & $-$77 31 53 &   8.01 & 0.40 &	0.29 &  0.93 &     ISO 168 & 12\\			    
64& 11 08 57.2 & $-$77 43 28 &   9.77 & 1.79 &	1.30 &	1.79 &     Denis 35 & 2\\			    
65& 11 08 58.6 & $-$77 33 57 &  10.94 & 0.78 &	0.28 &	1.04 &             & \\			     
66& 11 08 58.8 & $-$76 46 39 &  10.77 & 0.04 &0.29 & 0.84 &     ISO 170 & 12	\\		 
67& 11 09 03.8 & $-$77 00 51 &   8.09 & 0.32 &	0.35 & 	1.10 &     ISO 172 & 12	\\		 
68& 11 09 04.3 & $-$77 01 53 &  10.75 & 0.40 &	0.44 &	1.19 &     ISO 173 & 12	\\		 
69& 11 09 04.4 & $-$77 07 45 &   8.86 & 0.45 &	0.32 &	1.04 & 			  &  \\  
70& 11 09 08.2 & $-$76 18 14 &   6.04 & 0.58 &	0.33 &	0.69 & 			  &  \\
71& 11 09 08.5 & $-$76 49 13 &  10.42 & 0.42 &	0.64 &	0.90 &     ISO 177& 12\\		 
72& 11 09 11.4 & $-$76 32 50 &  10.75 & 1.11 &	0.96 &	1.86 &     Denis 36, NIR 2 & 2, 10\\		 
73& 11 09 12.0 & $-$77 39 06 &   9.91 & 0.56 &	0.71 &	1.82 &     Denis 37, ISO 179 & 2, 12\\		
74& 11 09 12.3 & $-$77 29 12 &   8.14 & 0.53 &	0.19 &	0.84 &     Sz 30, ISO 182 & 13, 12\\		    
75& 11 09 14.4 & $-$76 28 40 &  10.28 & 0.51 &	0.55 &	0.47 &     Hn 8	& 7\\		      
76& 11 09 15.4 & $-$76 21 58 &  10.60 & 0.12 &	0.19 &	0.69 & 			 & \\  
77& 11 09 18.3 & $-$76 27 58 &   8.34 & 0.35 &	0.28 &	1.01 &     CHXR 37, ISO 185 & 4, 12\\		
78& 11 09 18.5 & $-$77 47 40 &   7.09 & 0.72 & $-$0.14~~ & 0.23 &   ISO 187	& 12\\		   
79& 11 09 19.0 & $-$76 30 30 &   7.88 & 0.94 & 1.00 &	1.59 &     Hn 9, ISO 186	& 7, 12\\	  
80& 11 09 21.3 & $-$77 36 54 &  11.26 & 0.58 & 0.50 &	1.45 &  		& \\	     
81& 11 09 21.9 & $-$77 13 58 &   8.15 & 0.47 & 0.43 &	1.27 &     ISO 188	& 12 \\		  
82& 11 09 23.5 & $-$76 34 32 &   7.30 & 1.22 & 1.61 &	2.39 &     C1-6, ISO 189, NIR 10 & 9, 12, 10\\
83& 11 09 24.3 & $-$76 23 22 &   sat  &      &      &  0.91 &     VZ Cha	& 13\\
84& 11 09 25.9 & $-$77 26 25 &   8.63 & 0.71 & 0.58 &	1.59 &     ISO 190	& 12\\		    
85& 11 09 26.9 & $-$76 33 33 &   9.38 & 0.79 & 1.02 &	2.10 &     Denis 40, NIR 13	& 2, 10\\	
86& 11 09 27.3 & $-$77 36 52 &  10.40 & 0.42 & 0.41 &	1.04 &  			  \\  
87& 11 09 29.4 & $-$76 33 28 &   8.32 & 3.86 & 2.42 &	3.55 &    Denis 41, ChaINa2 \#1, ISO 192, NIR 15 & 2, 11, 12, 10\\	
88& 11 09 31.8 & $-$76 30 36 &  10.37 & 1.01 &	0.77 &	1.35 &    ISO 193 & 12 \\			      
89& 11 09 37.4 & $-$77 26 37 &  10.17 & 0.49 &	0.55 &	1.61 &    ISO 195 &  12 \\			      
90& 11 09 38.0 & $-$77 10 41 &   8.59 & 0.52 &	0.70 &	1.75 &    Denis 42, ISO 196 & 2, 12\\		  
91& 11 09 39.7 & $-$76 52 26 &  10.50 & 0.47 &	0.49 &	1.42 &    ISO 197  & 12 \\			      
92& 11 09 39.7 & $-$76 28 38 &   8.55 & 0.38 &	0.29 &	0.96 &    CHX 15b, ISO 198 & 5, 12\\		  
93& 11 09 41.8 & $-$76 34 59 &   8.45 & 1.18 &	1.51 &	2.50 &    C1-25, ISO 199, NIR 19 & 9, 12, 10\\	  
94& 11 09 42.1 & $-$76 51 11 &  10.78 & 0.72 &	0.73 &	1.64 &    & \\			      
95& 11 09 43.0 & $-$77 25 58 &   9.55 & 0.93 &	0.64 &	1.36 &    C7-1, ISO 200 & 9, 12\\		      
96& 11 09 45.1 & $-$77 40 32 &  10.03 & 0.44 &	0.45 &	1.00 &    ISO 201 & 12 \\			      
97& 11 09 46.2 & $-$76 34 47 &   9.20 & 0.76 &	0.72 &	1.37 &    Hn 10, ISO 204, NIR 24 & 7, 12, 10\\	  
98& 11 09 46.2 & $-$76 28 57 &   7.47 & 0.38 &	0.20 &	0.46 &    CHX 15a, ISO 205 & 5, 12\\		  
99& 11 09 46.8 & $-$76 43 54 &   9.51 & 0.63 &	0.58 &	1.11 &    C2-3, ISO 203 & 9, 12\\		      
100& 11 09 47.7 & $-$76 34 06 &  10.72 & 1.14 &	1.24 &	2.43 &    Denis 44, C1-22, NIR 25 & 2, 9, 10\\  
101& 11 09 47.9 & $-$77 26 30 &   8.58 & 1.72 &	0.82 &	1.62 &    B 43, ISO 207 & 1, 12\\		  
102& 11 09 49.3 & $-$77 31 20 &  10.62 & 0.45 &	0.38 &	1.02 &                   & \\				      
103& 11 09 49.9 & $-$76 36 49 &   6.85 &  	  &      &       &    HD 97300, ISO 2 & 211\\  
104& 11 09 50.6 & $-$77 45 48 &   8.10 & 1.25 &	0.60 &	1.38 &    ISO 214 &	12\\		     
105& 11 09 51.5 & $-$76 58 57 &   8.54 & 0.22 &	0.17 &	0.54 &    ISO 213 &	12 \\		     
106& 11 09 52.0 & $-$76 39 12 &  10.76 & 0.89 &	0.69 &	1.02 &    ISO 217 & 12 \\			     
107& 11 09 52.4 & $-$76 57 58 &   9.78 & 0.30 &	0.49 &	1.33 &    ISO 216 & 12\\			     
108& 11 09 53.3 & $-$77 45 40 &   7.92 & 0.16 &	0.42 &	1.17 &    ISO 222 & 12\\			     
109& 11 09 53.7 & $-$76 34 26 &   sat  &  	  & 1.44 &	2.16 &    HM 23, ISO 223, NIR 27  & 8, 12, 10\\	 
110& 11 09 53.9 & $-$76 29 25 &   8.22 & 0.88 &	0.74 &	1.35 &    Sz 33, CHX 15, ISO 224 & 13, 5, 12 \\
111& 11 09 55.0 & $-$76 32 40 &   8.05 & 1.20 &	1.73 &	2.61 &    C1-2, ISO 226, NIR 8 & 9, 12, 10\\	 
112& 11 09 56.7 & $-$77 18 26 &   9.33 & 0.54 &	0.75 &	1.87 &    Denis 47, ISO 227 & 2, 12 \\		 
113& 11 09 58.5 & $-$76 59 15 &   8.51 & 0.43 &	0.44 &	1.21 &    ISO 229 & 12, \\			    
114& 11 09 58.9 & $-$77 37 09 &   6.95 & 0.73 &	0.69 &	1.08 &    WX Cha, ISO 228 & 13, 12 \\		 
115& 11 10 00.0 & $-$77 26 33 &   8.98 & 0.94 &	0.59 &	1.60 &    ISO 230 & 12 \\			     
116& 11 10 00.7 & $-$76 34 59 &   sat  &  	  & 0.90 &	1.16 &    WW Cha, ISO 231, NIR 29 & 13, 12, 10\\	 
117& 11 10 04.1 & $-$76 33 28 &   8.27 & 0.93 &	0.85 & 	1.60 &    Hn 11, ISO 232, NIR 33	 & 7, 12, 10 \\ 
118& 11 10 04.9 & $-$76 35 47 &   8.56 & 0.54 &	0.38 &	1.00 &    GK-1, ISO 233, NIR 35 & 13, 12, 10\\	 
119& 11 10 07.1 & $-$76 29 38 &   7.86 & 0.62 &	0.45 &	0.96 &    WY Cha, ISO 234 & 13, 12\\		 
120& 11 10 08.8 & $-$77 27 50 &  10.09 & 1.09 &	0.90 &	1.49 &    ISO 235 & 12\\			     
121& 11 10 11.9 & $-$76 35 31 &   7.69 & 0.86 &	0.79 & 	1.53 &    Denis 48, ISO 237, NIR 45	& 2, 12, 10\\ 
122& 11 10 13.3 & $-$77 27 11 &   9.85 & 0.62 &	0.56 &	1.47 &    ISO 239 & 12\\			    
123& 11 10 30.5 & $-$77 36 06 &   9.59 & 0.19 &	0.33 &	1.09 &    ISO 243 & 12\\			    
124& 11 10 39.2 & $-$77 32 42 &   7.38 & 0.50 &	0.46 &	0.92 &    CHXR 47, ISO 251 & 4, 12\\		
\enddata
 
\tablecomments{Units of right ascension are hours, minutes, and seconds, and
units of declination are degrees, arcminutes, and arcseconds. Sources with
saturated detections on the SPIREX/Abu images are indicated by `sat'.}

\tablerefs{(1) \citet{bau84}, (2) \citet{cam98}, 
(3) \citet{com99}, (4) \citet{fei93},
(5) \citet{fekr89}, (6) \citet{gla79},
(7) \citet{hart93}, (8) \citet{heme73},
(9) \citet{hyl82}, (10) \citet{oas99},
(11) \citet{per99}, (12) \citet{per00},
(13) \citet{sch77}, (14) \citet{sche83}}

\end{deluxetable}                                             


\newpage

\begin{thebibliography}{}

\bibitem[Alcal\'a el al. (1995)]{alc95} Alcal\'a, J. M., Krautter, J., Schmitt,
J. H. M. M., Covino, E., Wichmann, R., \& Mundt, R.  1995, \aaps, 114, 109

\bibitem[Allen (1972)]{all72} Allen, D. A. 1972, \apjl, 172, L55

\bibitem[Appenzeller et al. (1983)]{app83} Appenzeller, I., Jankovics,
I., \& Krautter, J. 1983, \aap, 53, 291

\bibitem[Barsony et al. (1997)]{bar97} Barsony, M., Kenyon, S. J., Lada, E. A.,
Teuben, P. J. 1997, \apjs, 112, 109

\bibitem[Basri \& Batalha (1990)]{bas90} Basri, G., \& Batalha, C. 1990, 
\apj, 363, 654

\bibitem[Baud et al. (1984)]{bau84} Baud, B., Young, E., Beichman, C. A.,
Beintema, D. A., Emerson, J. P., Habing, H. J., Harris, S., \& Jennings, R. F.
1984, \apj, 278, L53

\bibitem[Bertout et al. (1988)]{ber88} Bertout, C., Basri, G., \& 
Bouvier, J. 1988, \apj, 330, 350

\bibitem[Bessell \& Brett (1988)]{bes88} Bessell, M. S., \& Brett, J. M. 
1988, \pasp, 100, 1134

\bibitem[Burrows et al. (1996)]{bur96} Burrows, C. J., Stapelfeldt, K. R.,
Watson, A. M., Krist, J. E., Ballester, G. E., Clarke, J. T., Crisp, D.,
Gallagher, J. S., III, Griffiths, R. E., Hester, J. J., Hoessel, J. G.,
Holtzman, J. A., Mould, J. R., Scowen, P. A., Trauger, J. T.,
\& Westphal, J. A. 1996, \apj, 473, 437

\bibitem[Cambr\'esy et al. (1998)]{cam98} Cambr\'esy, L., Copet, E., 
Epchtein, N., de Batz, B., Borsenberger, J., Fouqu\'e, P., 
Kimeswenger, S., \& Tiph\'ene, D.  1998, \aap, 338, 977     

\bibitem[Chiang \& Goldreich (1997)]{chi97} Chiang, E. I., \& Goldreich, P. 
1997, \apj, 490, 368

\bibitem[Comer\'on et al. (1993)]{com93} Comer\'on, F., Rieke, G. H., 
Burrows, A., \& Rieke, M. J. 1993, \apj, 416, 185

\bibitem[Comer\'on et al. (1999)]{com99} Comer\'on, F., Rieke, G. H., \&
Neuh\"auser, R. 1999, \aap, 343, 477       

\bibitem[Edwards et al.  (1993)]{edw93} Edwards, S., Strom, S. E., 
Hartigan, P., Strom, K. M.,
Hillenbrand, L. A., Herbst, W., Attridge, J., Merrill, K. M.,
Probst, R., \& Gatley, I. 1993, \aj, 106, 372

\bibitem[Elias (1978)]{el78} Elias, J. 1978, \apj, 224, 453

\bibitem[Feigelson et al. (1993)]{fei93} Feigelson, E. D., Casanova, S.,
Montmerle, T., \& Guibert, J. 1993, \apj, 416, 623 

\bibitem[Feigelson \& Kriss  (1989)]{fekr89} Feigelson, E. D., \& Kriss, 
G. A. 1989, \apj, 338, 262 

\bibitem[Fowler et al. (1998)]{fow98} Fowler, A. M., Sharp, N., Ball, W., 
Schinckel, A., Ashley, M. C., Boccas, M., Storey, J. W., DePoy, D. L., 
Martini, P., Harper, A., Marks, R. 1998, SPIE proceedings, 3354, 1170

\bibitem[Gauvin \& Strom (1992)]{gau92} Gauvin, L. S., \& Strom, K. M. 1992,
\apj, 385, 217

\bibitem[Ghez et al. (1997)]{ghe97} Ghez, A. M., White, R. J., \& Simon, M.
1997, \apj, 490, 353

\bibitem[Glass (1979)]{gla79} Glass, I. S. 1979, \mnras, 187, 305
        
\bibitem[G\'omez \& Kenyon (2001)]{gom01} G\'omez, M., \& Kenyon, S. J. 2001, 
\aj, 121, No. 2

\bibitem[Haisch et al. (2000)]{hai00} Haisch., K. E., Lada, E. A., \& Lada,
C. J. 2000, \aj, 120, 1396

\bibitem[Haisch,~Lada \& Lada (2001)]{hai01} Haisch., K. E., Lada, E. A., \& Lada,
C. J. 2001, \aj, in press

\bibitem[Hartigan (1993)]{hart93} Hartigan, P. 1993, \aj, 105, 1511  

\bibitem[Hartigan et al. (1990)]{hart90} Hartigan, P., Hartmann, L., 
Kenyon, S., Strom, S. E., \& Skrutskie, M. F. 1990, \apjl, 354, L25

\bibitem[Hartmann et al. (1998)]{har98} Hartmann, L., Calvet, N., 
Gullbring, E., \& D'Alessio, P. 1998, \apj, 495, 385

\bibitem[Hartmann \& Kenyon  (1990)]{hrt90} Hartmann, L., \& 
Kenyon, S. J. 1990, \apj, 349, 190

\bibitem[Hartmann \& Kenyon  (1996)]{har96} Hartmann, L., \& 
Kenyon, S. J. 1996, \araa, 34, 207

\bibitem[Henize \& Mendoza  (1973)]{heme73} Henize, K. G., \& Mendoza, E. E.
1973, \apj, 180, 115  

\bibitem[Huenemoerder et al. (1994)]{hue94} Huenemoerder, D. P., Lawson, W. A., 
\& Feigelson, E. D. 1994, \mnras, 271, 967

\bibitem[Hyland  (1980)]{hyl80} Hyland, A. R. 1980, in 
{\it Infrared Astronomy, IAU Symp.  No. 96}, edited by 
C. D. Wynn-Williams \& D. P. Cruikshank, Reidel, Dordrecht, p. 125

\bibitem[Hyland et al. (1982)]{hyl82} Hyland, A. R., Jones, T. J., 
\& Mitchell, R. M.  1982, \mnras, 201, 1095
 
\bibitem[Kenyon \& Hartmann  (1987)]{ken87} Kenyon, S. J., \& Hartmann, L.
1987, \apj, 323, 714

\bibitem[Kenyon \& Hartmann  (1995)]{kh95} Kenyon, S. J., \& Hartmann, L.
1995, \apjs, 101, 117

\bibitem[Kenyon et al. (1998)]{ken98} Kenyon, S. J., Lada, E. A., \& Barsony, M.
1998, \aj, 115, 252

\bibitem[Kenyon et al. (1996)]{ken96} Kenyon, S. J., Yi, I., \& Hartmann, L.
1996, \apj, 462, 439

\bibitem[Koerner et al. (1993)]{koe93} Koerner, D. W., Sargent, A. I., 
\& Beckwith, S. V. W. 1993, \apjl, 408, 93 

\bibitem[Krist et al. (2000)]{kri00} Krist, J. E., Stapelfeldt, K. R., 
M\'enard, F., Padgett, D. L., \& Burrows, C. J. 2000, \apj, 538, 793

\bibitem[Lada  (1999)]{lad99} Lada, C. J. 1999,
in {\it The Physics of Star Formation and Early Stellar Evolution,}
edited by C. J. Lada \& N. Kylafis, Dordrecht, Kluwer, p. 143

\bibitem[Lada et al. (2000)]{lad00} Lada, C. J.,
Muench, A. A., Haisch, K. E. Jr., Lada, E. A., 
Alves, J. F., Tollestrup, E. V., Willner, S. P. 2000, 
\aj, 120, 3162

\bibitem[Lawson et al. (1996)]{law96}Lawson, W. A., Feigelson, E. D., 
\& Huenemoerder, D. P. 1996, \mnras, 280, 1071

\bibitem[Lynden-Bell \& Pringle  (1974)]{lyn74} Lynden-Bell, D., \& 
Pringle, J. E. 1974, \mnras, 168, 603

\bibitem[Nordh et al. (1996)]{nor96} Nordh, L., Olofsson, G., Abergel, A., 
Andre, P., Blommaert, J., Bontemps, S., Boulanger, F.,
Burgdorf, M., Cesarsky, C. J., Cesarsky, D.,
Copet, E., Davies, J., Falgarone, E., Huldtgren, M.,
Kaas, A. A., Lagache, G., Montmerle, T.,
Perault, M., Persi, P., Prusti, T., Puget, J. L.,
\& Sibille, F.  1996, \aap, 315, L185
 
\bibitem[Oasa et al. (1999)]{oas99} Oasa, Y., Tamura, M., \& Lawson, W. A.
1999, \apj, 526, 336    

\bibitem[Olofsson et al. (1999)]{olo99} 
Olofsson, G., Huldtgren, M., Kaas, A. A., Bontemps, S., Nordh, L., Abergel, 
A., Andr\'e, P., Boulanger, F., Burgdorf, M., Casali, M. M., Cesarsky, C. J., 
Davies, J., Falgarone, E., Montmerle, T., Perault, M., Persi, P., Prusti, T., 
Puget, J. L., \& Sibille, F. 2000, \aap, 357, 1020                                                                   
\bibitem[Persi et al. (1999)]{per99} Persi, P., Marenzini,  A. R., Kaas, A. A.,
Olofsson, G.,  Nordh, L., \& Roth, M. 1999, \aj, 117, 439

\bibitem[Persi et al. (2000)]{per00} Persi, P., Marenzini, A. R., Olofsson, G.,
Kaas, A. A., Nordh, L., Huldtgren, M., Abergel, A., Andr\'e, P., Bontemps, S.,
Boulanger, F., Burgdorf, M., Casali, M. M., Cesarsky, C. J., Copet, E., Davies, J.,
Falgarone, E., Montmerle, T., Perault, M., Prusti, T., Puget, J. L., \&
Sibille, F. 2000, \aap, 357, 219                                                                  
\bibitem[Press et al. (1992)]{pre92} Press, W. H., Flannery, B. P., 
Teukolsky, S. A., \& Vetterling, W. T.  1992, {\it Numerical Recipes, 
The Art of Scientific Computing,} Cambridge, Cambridge

\bibitem[Prusti,~Whittet \& Wesselius (1992)]{pru92} Prusti, T., 
Whittet, D. C. B., \& Wesselius, P. R. 1992, \mnras, 254, 361

\bibitem[Rucinski (1985)]{ruc85} Rucinski, S. M. 1985, \aj, 90, 2321.

\bibitem[Rydgren \& Zak  (1987)]{ryd87} Rydgren, A. E., \& Zak, D. S. 1987, 
\pasp, 99, 141

\bibitem[Schwartz  (1977)]{sch77} Schwartz, R. D. 1977, \apjs, 35, 161  

\bibitem[Schwartz \& Henize (1983))]{sche83} Schwartz, R. D., \& Henize, K. G.
1983, \aj, 88, 1665 

\bibitem[Valenti et al. (1993)]{val93} Valenti, J. A., Basri, G., \& 
Johns, C. M. 1993, \aj, 106, 2024

\bibitem[Walter (1992)]{wal92} Walter, F. M. 1992, \aj, 104, 758 

\bibitem[Wilking \& Lada (1983)]{wil83} Wilking, B. A., \& Lada, C. J. 
1983, \apj, 274, 698

\bibitem[Wilner et al. (1996)]{wil96} Wilner, D. J., Ho, P. T. P., \&
Rodriguez, L. F. 1996, \apj, 470, L117

\bibitem[Wilner et al. (2000)]{wil00} Wilner, D. J., Myers, P. C.,
Mardones, D., \& Tafalla, M. 2000, \apj, 544, L69

\end{thebibliography}
\end{document}